\documentclass{JAC2003}

\usepackage{graphicx}
\usepackage{booktabs}
\usepackage{color}
\usepackage{subfigure}
\usepackage{multirow}
\definecolor{red}{rgb}{1,0,0}
\definecolor{blue}{rgb}{0,0,1}

\begin{document}

\title{BENCHMARKING ELECTRON-CLOUD SIMULATIONS AND PRESSURE MEASUREMENTS AT THE LHC}

\author{O. Dom\'{i}nguez\thanks{Email: octavio.dominguez@cern.ch. Also at EPFL, 
LPAP, CH-1015 Lausanne, Switzerland} and F.~Zimmermann, CERN, Geneva, Switzerland}

\maketitle

\begin{abstract}

During the beam commissioning of the Large Hadron Collider (LHC)~\cite{lhc-dr,lhc-rev} with 150, 75, 50 and 25-ns bunch spacing, important electron-cloud effects, like pressure rise, cryogenic heat load, beam instabilities or emittance growth, were observed.
A method has been developed to infer different key beam-pipe surface parameters by benchmarking simulations and pressure rise observed in the machine. This method allows us to monitor the scrubbing process (i.e.~the reduction of the secondary emission yield as a function of time) in the regions where the vacuum-pressure gauges are located, in order to decide on the most appropriate strategies for machine operation. In this paper we present the methodology and first results from applying this technique to the LHC.

\end{abstract}

\section{Introduction}

Since almost 15 years photoemission and secondary emission had been predicted 
to build up an electron cloud inside the LHC beam pipe \cite{FZ1}, similar to the photo-electron instability in positron storage rings \cite{satoh,ohmi,furmanl}.
The possibility of ``beam-induced multipacting'' at the LHC had been suggested even earlier \cite{Groebner} extrapolating from observations with bunched beams at the ISR in the 1970s \cite{oswaldisr}. 
The electron cloud, at sufficiently high density, 
can cause both single and coupled-bunch instabilities of the proton beam \cite{FZ1,FZ2}, 
give rise to incoherent beam losses or emittance growth \cite{EBGFFZ,RHIC2}, heat the vacuum chamber (and subsequently provoke a quench in superconducting magnets),  
or lead to a vacuum pressure increase by several orders of magnitude due to electron stimulated desorption \cite{JMJ1}. 
All these effects eventually lead to luminosity limitations. Specifically, electron-cloud induced pressure rises 
have been one of the main  performance limitations for some accelerators \cite{RHIC2}.
From 1999 onward electron-cloud effects have been seen with LHC-type beams first in the SPS, then in the PS,
and finally, since 2010, as expected, in the LHC itself.  
During the early LHC beam commissioning with 150, 75 and 50-ns bunch spacing important electron-cloud effects, 
such as pressure rise, cryogenic heat load, beam instabilities, beam loss and emittance growth, were observed \cite{JMJ2,GA1,GR1}. 
Several exploratory studies at the design bunch spacing of 25 ns were performed during 2011~\cite{GR2}. 

The LHC mitigation strategy against electron cloud includes a sawtooth pattern on the horizontally outer side 
of the so-called beam screen inside the cold arcs, a shield mounted on top of the beam-screen pumping slots blocking the 
direct path of electrons onto the cold bore of the magnets, 
NEG coating for all the warm sections of the machine, installation of 
solenoid windings in field-free portions of the interaction region, and, last not least, beam scrubbing, i.e.~the reduction of the Secondary 
Emission Yield (SEY) with increasing electron dose hitting the surface, i.e.~as a result of the electron cloud itself. 
Beam scrubbing represents the ultimate mitigation of electron-cloud effects of the LHC, 
and it is considered necessary to achieve nominal LHC performance \cite{OB1}. 

At injection energy (450 GeV), the pressure inside the vacuum beam pipe affects the speed of the 
electron-cloud build up, since the initial electrons are produced by gas ionization. 
However, if there is noticeable multipacting the rate of primary electrons does not significantly influence the final value 
of the saturated electron density, 
which is then determined by secondary emission (multipacting) and by the space-charge field of the electron cloud itself. 
In such case larger vacuum pressures just make the electron density reach its equilibrium value faster. 
This is due to the fact that the energy spectrum of electrons hitting the wall is insensitive to the pressure~\cite{UI1}. 

Nevertheless,
in order to infer the best estimates of the beam-pipe characteristics, 
the steady-state vacuum pressure of the machine for each stage of observation has to be introduced 
as a simulation input parameter, in order to correctly account for the multiturn nature of the pressure evolution in a circular accelerator like the LHC. 
This is due to the fact that the time constant of the vacuum evolution is much longer than the revolution period,  
while the electron-cloud build-up simulations typically model only a fraction of a turn.  
According to the vacuum-gauge measurements, a steady-state pressure is normally established a few minutes after injecting the last bunch train for a given configuration.

\begin{figure}[htb]
  \centering
  \includegraphics[width=80mm]{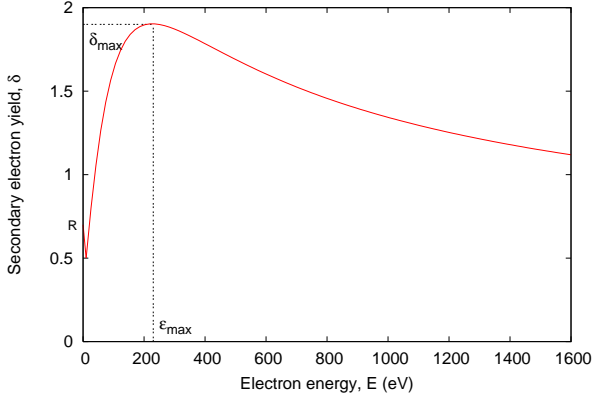}
  \caption{Secondary emission yield for perpendicular incidence as a function of primary electron energy, defining the parameters $\delta_{\textrm {max}}$, $\varepsilon_{\textrm {max}}$ and \textit{R}.}
  \label{SEYvsE}
\end{figure}

Since dedicated in-situ measurements of the LHC electron-cloud density and the LHC vacuum-chamber surface properties are not available 
we are developing a method to determine the actual surface properties of the vacuum chamber related to secondary emission
and to the electron-cloud build up ($\delta_{\textrm {max}}$, $\varepsilon_{\textrm {max}}$ and \textit{R}~\cite{cimino}; see Fig.~\ref{SEYvsE} for a graphical definition of these three  quantities),  
and their evolution in time, 
based on benchmarking computer simulations of the electron flux on the chamber surface using the ECLOUD code 
against pressure measurements for different beam characteristics. 
This new method allows monitoring the effectiveness of LHC ``scrubbing runs'' 
and provides snapshots of the surface conditions around the LHC ring.

\section{Methodology}

The pressure increase due to an electron cloud can be related with the electrons hitting the chamber wall as
\begin{equation}
  \Delta P=kT\frac{\int{\eta_e(E)\phi_e(E)dE}}{S_{eff}}\,,
\end{equation}
where $k$ denotes Boltzmann's constant, $T$ the temperature, $S_{eff}$ the pumping speed, $\eta_e$ the 
electronic desorption coefficient and $\phi_e$ the flux of electrons hitting the chamber wall. 
The quantities $\eta_e$ and $S_{eff}$ cannot be introduced in the present electron cloud simulation codes,
but assuming that the pressure increase is proportional to the electron flux hitting the chamber wall, 
pressure measurements for different bunch train configurations (e.g.~with changing spacing 
between trains or with a varying number of trains injected into the machine) can be benchmarked against simulations 
by comparing ratios of observed pressure increases and of simulated electron fluxes at the wall, respectively. 
The idea of the benchmarking using ratios goes back to an earlier study for the SPS (serving as LHC injector) 
where the electron-cloud flux could be measured directly ~\cite{DS}. In the LHC case, no electron-cloud monitor is available, 
but instead the measured increase in the vacuum pressure is taken to be a reliable indicator 
proportional to the electron flux on the wall. 

We face a four-parameter problem. 
The steps followed in the benchmarking are the following: 

(1) We fix two of the parameters, namely the pressure 
(using the measured value) and $\varepsilon_{\textrm {max}}$ (set to 230 eV, which seems to be a good first estimate 
according to past surface measurements and some previous simulation benchmarking~\footnote{Several studies (e.g.~\cite{CYV}) reveal an evolution of the value of $\varepsilon_{\textrm {max}}$ with the scrubbing process. 
This evolution depends on the scrubbing technique (either using an electron gun or a real beam) 
and several parameters such as the roughness of the surface, the previous surface treatment, the electrons energy, etc. 
Simulations depend indeed on this parameter. 
Further investigation is currently ongoing to infer its evolution in the LHC.}).

(2) We simulate the electron cloud build up for different bunch configurations using the ECLOUD code, 
scanning the other two parameters, $\delta_{\textrm {max}}$ and \textit{R}, in steps of 0.1 and 0.05 respectively. 
Smaller steps introduce statistical noise which needs to be controlled by smoothing techniques.

(3) For each bunch configuration we plot the simulated electron flux $\phi_{i}$ 
above a 2D grid spanned by $\delta_{\textrm {max}}$ and \textit{R}. 

(4) We fit the flux simulated on the grid to a third order polynomial 
and then form the ratio of simulated fluxes (that is, dividing the polynomials) 
for two different bunch configurations [the fluxes and not their ratio are fitted 
in order to suppress the effect of statistical fluctuations]. 

\begin{figure}[htb]
  \centering
  \includegraphics[width=80mm]{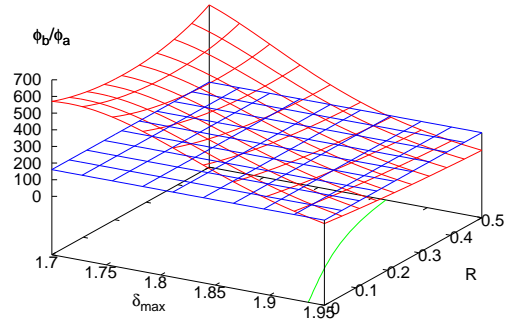}
  \caption{Example of a 3D surface of simulated fluxes for 
the case $\phi_b/\phi_a$ (red) cut by a plane surface (blue) at the value equal to the ratio of the corresponding measured pressures ($P_b/P_a=161.0$). The bottom plane shows the contour of the intersection between both surfaces (green).}
  \label{3Dgrid}
\end{figure}

(5) Comparing the latter ratio with the experimental ratio 
of measured pressure increases yields a curve in the $\delta_{\textrm {max}}$-\textit{R} plane 
(see Fig.~\ref{3Dgrid}). Different configurations yield different curves in that plane. 

(6) If the measurements contain sufficient information and the simulation 
model is reasonably accurate we expect to obtain a unique intersection between lines 
corresponding to different bunch configurations. This crossing point then
defines the solution for $\delta_{\textrm {max}}$ and \textit{R}.

We apply this methodology for certain LHC regions in which pressure gauges and vacuum pumps are located. 
Normally these are installed in short beam-pipe modules made from copper-coated stainless steel mounted between two NEG coated pipes (7 m long each), with a good pumping and (after activation) low secondary emission yield, so that we may assume that the pressure rise measured at a gauge is the result of the electron cloud produced exclusively within the gauge's vacuum module. 
The module vacuum chamber is round with 80 mm diameter.
In the following we only present results for one ionization gauge. Results for other gauges look similar.

\section{Results}

Until now we have processed 4 sets of measurements obtained
during the conditioning of the machine through beam scrubbing. 
All of these have been recorded at a beam energy of 450 GeV with either 50-ns or 25-ns bunch spacing for the first 3 sets and the last set, respectively. 

We have used two kinds of beam configurations, with varying spacing between successive bunch trains (also called ``batches'') 
and varying number of batches, respectively.  
Table~\ref{SimParam} lists the parameters for the three sets of measurements with 50 ns bunch spacing.

\begin{table}[hbt] 
  \centering
  \caption{Parameters used in the simulations for the different sets of measurements with 50 ns bunch spacing. In sets 2 and 3 there is an additional space of 225 ns between the two (set 2) or three trains (set 3) of 36 bunches injected simultaneously.}
  \vspace{0.3cm}
  \begin{tabular}{l|c|c|c}
    \hline
                                   & \textbf{Set 1}        & \textbf{Set 2}                  & \textbf{Set 3}                \\
    \hline
    \textbf{\#} of bunches         & \multirow{2}{*}{36}   &  \multirow{2}{*}{72 (2x36)}     &  \multirow{2}{*}{108 (3x36)}  \\
     per batch                     &                       &                                 &                               \\
    \hline 
    \textbf{\#} of batches         &   1 - 5               &    1 - 14                       &   1 - 12                      \\
    \hline
    Batch spacing                  & \multirow{2}{*}{2.0 - 6.0} &    \multirow{2}{*}{4.850}  &   \multirow{2}{*}{0.925}      \\
    ($\mu$s)                       &                            &                            &                               \\ 
    \hline
    Av. bunch                      & \multirow{3}{*}{1.1}  &  \multirow{3}{*}{1.21}          &  \multirow{3}{*}{1.15}        \\ 
    population                     &                       &                                 &                               \\
    ($10^{11}$ ppb)                &                       &                                 &                               \\ 
    \hline
  \end{tabular}
  \label{SimParam}
\end{table}

\begin{figure}[htb]
  \centering
  \includegraphics[width=80mm]{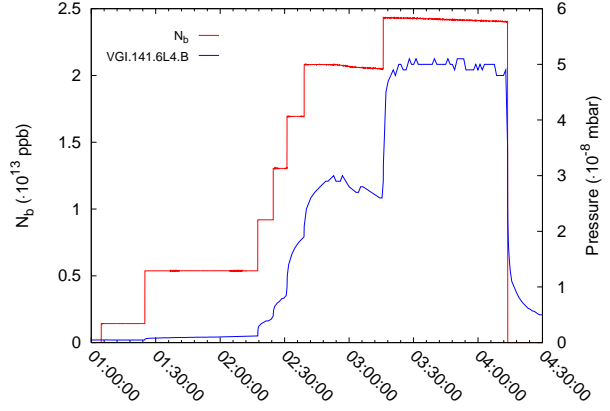}
  \caption{Beam intensity and pressure at the gauge VGI.141.6L4.B during the first experiment on 6 April 2011. Every step
in the beam intensity (red curve) indicates the injection of a new batch.}
  \label{6April1stexp}
\end{figure}

At the beginning of a scrubbing run in April 2011,  
two experiments were carried out (both corresponding to set 1 listed above). 
In the first we injected batches in pairs with varying 
batch spacing (6 $\mu$s, 4 $\mu$s and 2 $\mu$s). Each pair of batches was separated by 11.5 $\mu$s 
(a time considered long enough to clear any electron cloud).
Figure \ref{6April1stexp} shows the pressure increases
observed during this first experiment, including an additional first, shorter 12-bunch batch introduced for machine-protection reasons (and where no pressure increase can be appreciated). 
In the second experiment we injected an increasing number of batches 
at a batch-to-batch distance of 2.125 $\mu$s (up to 5). 

\begin{figure}[htb]
  \centering
  \includegraphics[width=80mm]{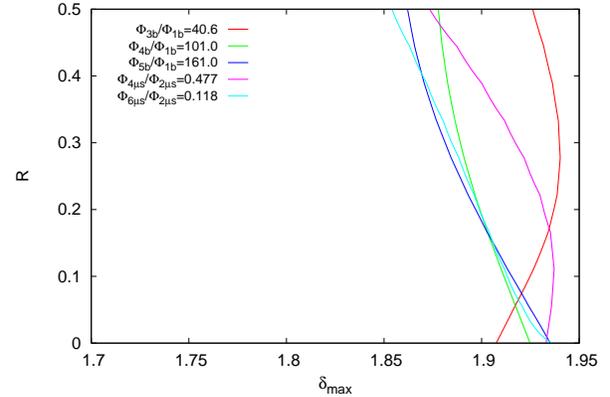}
  \caption{Combinations of $\delta_{\textrm {max}}$-\textit{R} values characterizing the chamber surface, obtained by
benchmarking ratios of observed pressure increases against ratios of simulated electron fluxes, 
for measurements on 6 April 2011.}
  \label{6April}
\end{figure}

Figure~\ref{6April} depicts the results obtained for both experiments.
We could conclude that the solution is around  $\delta_{\textrm {max}}=1.9$ and $R=0.2$.
We have to take into account that there are large 
uncertainties in the measured pressure values as well as in the estimated 
bunch population. According to simulations, such uncertainties 
can lead to a mismatch between lines and prevent a single 
unique intersection, as seen for this example. 
The value of $\delta_{\textrm {max}}=1.9$ is in agreement with an estimate from the CERN vacuum group,
which expected an initial value between 1.6 and 1.9 \cite{NH,CS}.
In addition, the value $R=0.2$ is in agreement with several high precision measurements, both recent (e.g.~\cite{ONERA}) and old (e.g.~\cite{K1}).

\begin{figure}[htb]
  \centering
  \includegraphics[width=70mm]{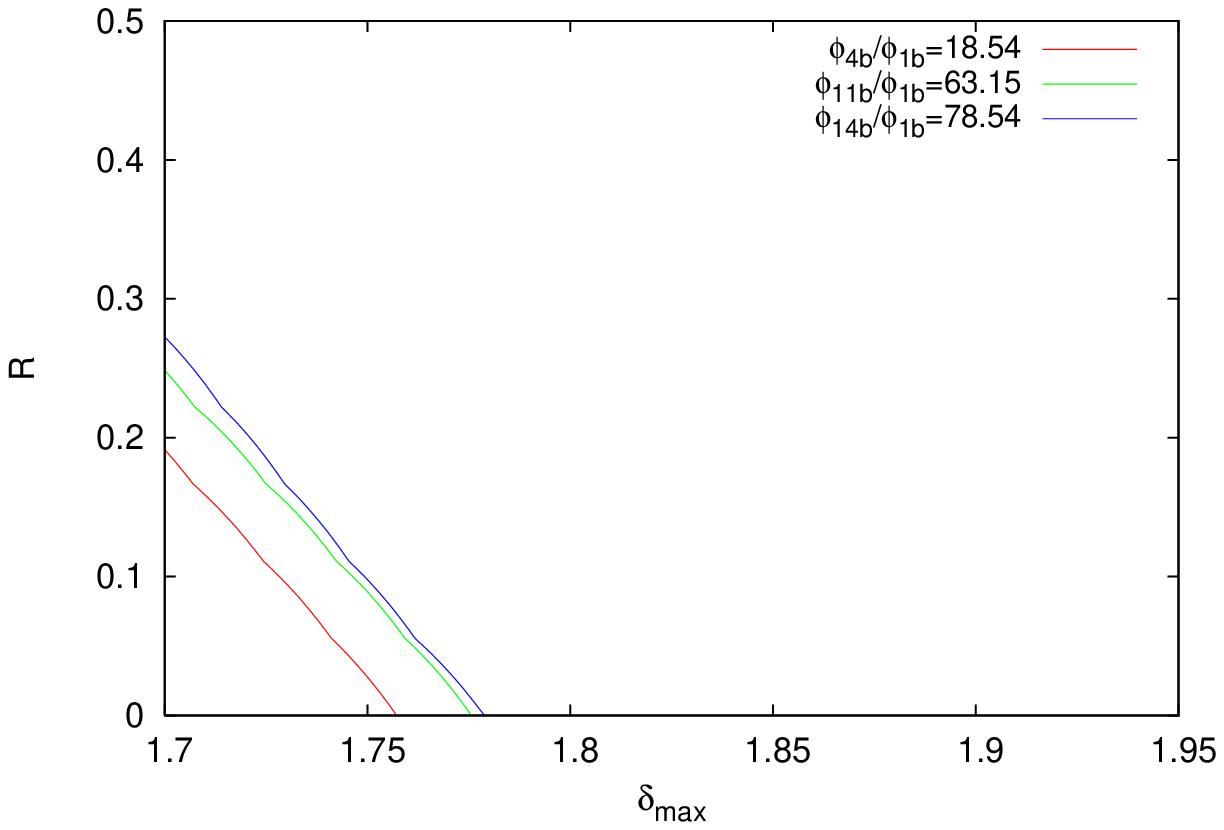}
  \includegraphics[width=70mm]{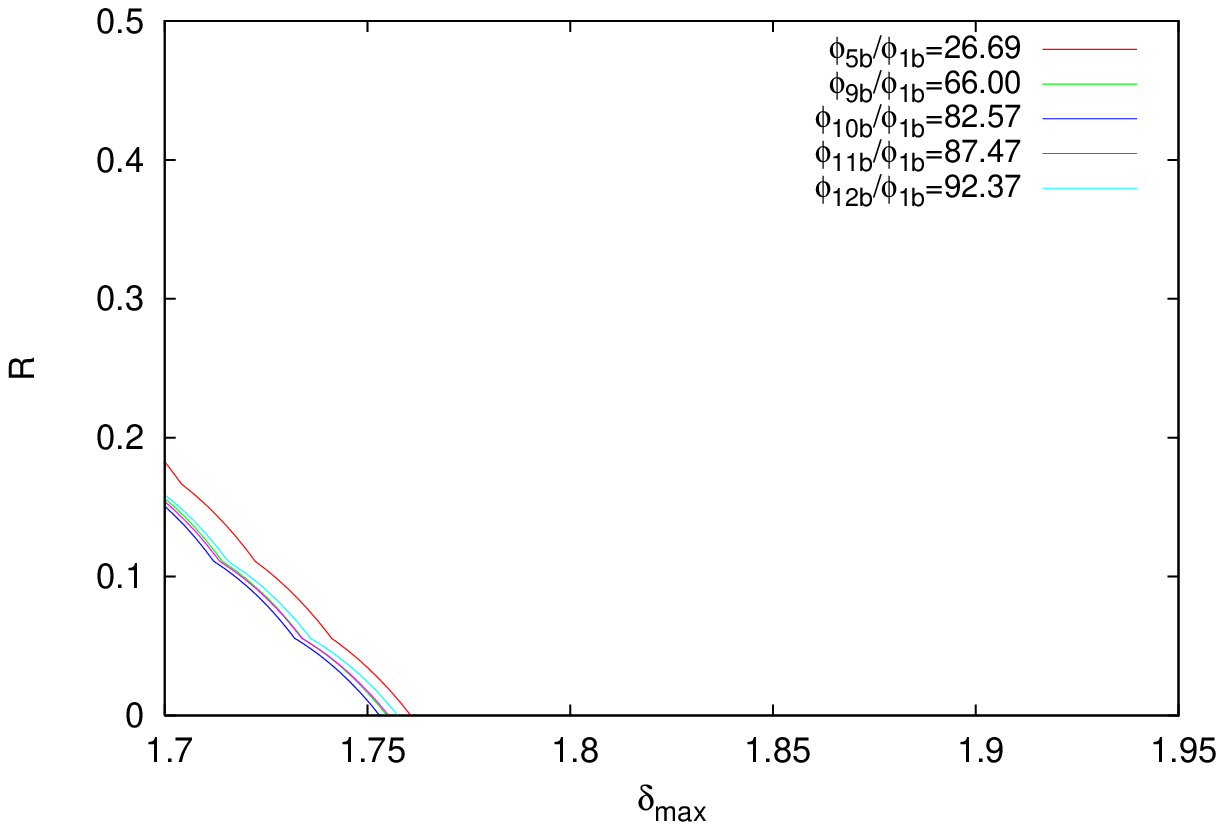}
  \caption{Combinations of $\delta_{\textrm {max}}$-\textit{R} values characterizing the chamber surface, obtained by
benchmarking ratios of observed pressure increases against ratios of simulated electron fluxes, 
for LHC measurements taken on 11 April (top) and 19 May 2011 (bottom).}
  \label{Parallel}
\vspace{-0.2cm}
\end{figure}

After a few days of surface conditioning, 
double batches of 36 bunches each separated by 225 ns were injected at a distance of 4.85 $\mu$s (up to 14). This corresponds to the set 2 of experimental data. 
A similar experiment (set 3) took place in mid May 2011 but using triple batches instead, again separated by 225 ns, 
at a distance of 925 ns (up to 12). 
Figure~\ref{Parallel} shows the results obtained in these cases.
It is worth noting that for these last two cases we observe parallel lines instead of a clear intersection between the lines.
This is due to the loss of sensitivity to the effect of the 225 ns gap between 36-bunch batches, that appears when the double (or triple) batches are injected together instead of one after the other. 
Indeed the lines should be identical under some plausible simplifying assumptions.  
The conclusion is that it is necessary, during the same experiment, to take two sets of measurements with different batch spacings, in order to obtain lines of different slope which uniquely intersect and yield the desired parameter information. 

\begin{figure}[htb]
  \centering
  \includegraphics[width=80mm]{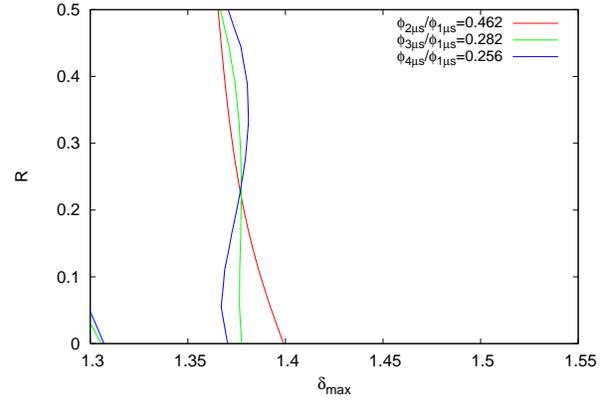}
  \caption{Combinations of $\delta_{\textrm {max}}$-\textit{R} values characterizing the chamber surface, obtained by
benchmarking ratios of observed pressure increases against ratios of simulated electron fluxes, 
for LHC measurements taken on 25 October 2011. In this case a value $\varepsilon_{\textrm {max}}=260$~eV has been assumed, as it gives a better fit to the data.  
This could be a sign of variation of $\varepsilon_{\textrm {max}}$ during the scrubbing.}
  \label{plot25ns}
\end{figure}

A new measurement, with varying spacing between batches,  
has been carried out at the end of the 2011's proton run (end October). 
On this occasion the bunch spacing was reduced to 25 ns. 
Table~\ref{25ns} shows the parameters used in this case and Fig.~\ref{plot25ns} depicts the result obtained for this experiment.

\begin{table}[hbt] 
  \centering
  \caption{Parameters used in the simulations for the measurements with 25 ns bunch spacing.}
  \vspace{0.2cm}
  \begin{tabular}{l|c}
    \hline
                                   & \textbf{Set 4}                      \\
    \hline
    \textbf{\#} of bunches         & \multirow{2}{*}{72}                 \\ 
    per batch                      &                                     \\
    \hline
    \textbf{\#} of batches         &  2                                  \\ 
    \hline  
    batch spacing ($\mu$s)         &   1.0, 2.0, 3.0, 4.0                \\ 
    \hline
    bunch population               & \multirow{2}{*}{1.1, 1.0, 1.0, 1.1} \\
    ($10^{11}$ ppb)           &                                     \\
    \hline
  \end{tabular}
  \label{25ns}
\end{table}

\begin{figure}[htb]
  \centering
  \includegraphics[width=80mm]{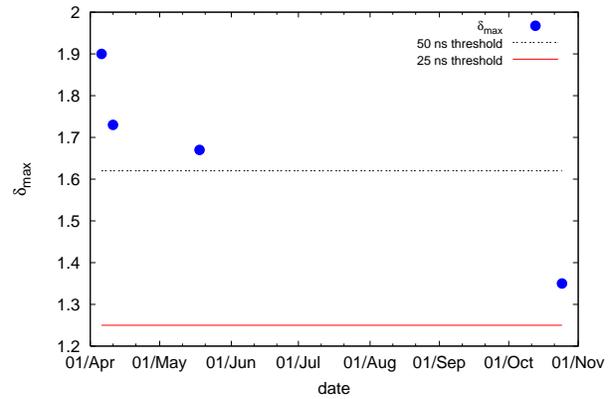}
  \caption{Approximate time evolution (from April to October 2011) of $\delta_{\textrm {max}}$ in the uncoated warm straight sections (chamber radius 40 mm).
The lowest values reachable 
with scrubbing corresponds to the 25-ns multipacting threshold,
for a reflectivity $R$ of 0.2.  
The first injection of a beam with 25-ns bunch spacing beam took place on 29 June 2011.
}
  \label{Summary}
\end{figure}

Although we are not yet able to extract a unique value for $\delta_{\textrm {max}}$ and \textit{R}, 
we can clearly see evidence for conditioning, as the solution for later cases tends towards lower $\delta_{\textrm {max}}$ values.
Figure~\ref{Summary} summarizes the approximate time evolution of $\delta_{\textrm {max}}$ in the ``warm-warm" transition regions where pressure gauges are located. 

We can see that the evolution of the conditioning for 50-ns and 25-ns beams looks as expected, with $\delta_{\rm max}$ approaching the simulated multipacting thresholds for 50-ns or 25-ns bunch spacing (both also indicated in the figure), respectively, as asymptotic limit.
All these facts instill some confidence in the method and support its potential use as a tool 
for monitoring the surface conditioning through beam scrubbing.  

The evolution in the uncoated straight sections goes from an initial value of $\delta_{\textrm {max}}\approx 1.9$ at the beginning of the scrubbing run in April 2011 to
$\delta_{\textrm {max}}\approx 1.35$ when the experiments with 25 ns were carried out.
The points shown for 11 April and 19 May 2011 have been obtained by assuming the value of an average line in Fig.~\ref{Parallel} at $R=0.2$. 

\section{Conclusions}

In 2010 and 2011 first electron-cloud effects have been observed with proton beams in the LHC. 
Rapid surface conditioning has allowed reducing the bunch spacing for nominal operation from 150 ns 
over 75 ns down to 50 ns without any significant perturbation from electron cloud. 

Thanks to the benchmarking of vacuum observations against simulations described in this paper, 
we have been able to monitor the evolution of $\delta_{\textrm {max}}$ during machine conditioning 
in the warm straight sections of the LHC. 
The observable considered is the pressure increase resulting from the electron cloud, 
which is taken to be proportional to the electron flux impinging on the vacuum chamber walls. 
Namely, by benchmarking the ratios of experimental pressures and of simulated electron fluxes 
for different beam configurations (e.g., for varying spacing between bunch trains or varying 
number of batches) we can then pin down the value of the maximum secondary emission yield as well as 
the reflection probability for low-energy electrons.  
Applying this method to each of the different measurement sets available so far provides clear evidence for surface conditioning in the uncoated warm regions of the LHC, from an initial maximum 
secondary emission yield of about 1.9 down to about 1.35, with $R\approx 0.2$, as can be seen in Fig.~\ref{Summary}.

In order to reach the design LHC bunch spacing of 25 ns in physics operation, 
further conditioning of the secondary emission yield is still required. According to some estimates \cite{GR3}, 
approximately 2 weeks of machine time would be required to achieve these values, since the 
scrubbing effect reduces with decreasing $\delta_{\textrm{max}}$.

\section{Acknowledgements}

The authors would like to thank 
G.~Arduini, V.~Baglin, G.~Bregliozzi, G.~Iadarola, G.~Lanza, E.~M\'{e}tral, 
G.~Rumolo, M.~Taborelli and C.~Yin-Vallgren for relevant and always enriching discussions and experimental
data as well as all the machine operators for the support they provided with the machine operation during the experiments.


\end{document}